\documentclass[
  journal=largetwo,
  manuscript=article-type,
  year=2024,
  volume=1,
]{cup-journal}

\usepackage{amsmath}
\usepackage[nopatch]{microtype}
\usepackage{booktabs}
\usepackage{amssymb}
\usepackage{hyperref}
\usepackage{soul} 

\newcommand*{\aap}{AAp}
\newcommand*{\apjl}{ApJL}

\newcommand*{\apjs}{ApJS}

\title{Adaptive particle refinement for compressible smoothed particle hydrodynamics}

\author{Rebecca Nealon}
\affiliation{Centre for Exoplanets and Habitability, University of Warwick, Gibbet Hill Road, CV4 7AL Coventry, UK}
\alsoaffiliation{Department of Physics, University of Warwick, Gibbet Hill Road, CV4 7AL Coventry, UK}
\email[R. Nealon]{rebecca.nealon@warwick.ac.uk}

\author{Daniel J. Price}
\affiliation{School of Physics and Astronomy, Monash University, Clayton, Vic 3800, Australia}


\begin{document}
\emergencystretch 3em

\begin{abstract}
We introduce adaptive particle refinement for compressible smoothed particle hydrodynamics (SPH). SPH calculations have the natural advantage that resolution follows mass, but this is not always optimal. Our implementation allows the user to specify local regions of the simulation that can be more highly resolved. We test our implementation on practical applications including a circumbinary disc, a planet embedded in a disc and a flyby. By comparing with equivalent globally high resolution calculations we show that our method is accurate and fast, with errors in the mass accreted onto sinks of less than $9\%$ and speed ups of $1.07 - 6.62\times$ for the examples shown. Our method is adaptable and easily extendable, for example with multiple refinement regions or derefinement.
\end{abstract}

This paper considers the best way to locally adapt resolution in the simulation method known as smoothed particle hydrodynamics \citep[SPH,][]{Lucy:1977lr2,gingold_monaghan_1977}. Because resolution follows mass in SPH, the best resolved region often corresponds to the region of interest. However, this may be inefficient, as high densities correlate with short timesteps which are expensive. A potential solution is  Adaptive Particle Refinement (APR): splitting and merging of particles. This method allows multiple resolutions to co-exist in the same simulation \citep{Monaghan:1988lk,Meglicki:1993hj,Kitsionas:2002bh,Kitsionas:2007ih,Vacondio:2013wd,Lopez:2013nj,Barcarolo:2014vu,Chiron:2018hq,Gao:2022wd}. APR is conceptually similar to Adaptive Mesh Refinement \citep[AMR,][]{Berger:1989bh,Truelove:1997jw} commonly used in simulations employing Eulerian meshes. 

Although less widely exploited than mesh refinement, the fundamentals of particle refinement and its applications are well developed. \citet{Monaghan:1988lk} and \citet{Meglicki:1993hj} first applied particle splitting in SPH. They merged particles in their simulations to create fewer, more massive particles when the density in a cell exceeded a critical density threshold. \citet{Meglicki:1993hj} additionally included particle splitting in low density regions to increase the local resolution. In the context of cosmological simulations, based on the method outlined in \citet{Porter:1985vs}, \citet{Katz:1993bu} (and later \citealt{Navarro:1994ku}) scaled the mass of the particles in spherical, nested layers throughout the computational domain such that the innermost layer had the smallest mass and thus highest resolution. \citet{Thacker:2000vg} found that there was a small amount of noise generated at the boundary between the mass layers but that it did not affect the structures identified in their simulations.  \citet{Bertschinger:2001aw} produced a method to accurately initialise simulations with multiple masses for a Gaussian random field. \citet{Klypin:2001bf} demonstrated numerical convergence with this simulation style. This `multiple mass' or `zoom' method, the approach of nested regions is a core feature of cosmological simulations \citep[e.g.][]{Springel:2005wk}.

\citet{Borve:2001oe} implemented an alternative multi-resolution scheme they called Regularised-SPH (RSPH). In this method, the smoothing length $h$ was set to a piece-wise constant in steps of $2$ with the contribution from neighbouring particles interpolated using a grid. In its most modern invocation \citep{Borve:2005as}, this method used auxiliary particles that exist at the boundary between different $h$ regions and which are passively evolved. \citet{Borve:2006bq} tested RSPH on a multidimensional MHD shock, showing excellent shock capturing properties. \citet{Borve:2009cw} also successfully applied RSPH to planet-disc interactions in two dimensions.

\citet{Kitsionas:2002bh} considered the collapse of clumps in self gravitating filaments. They increased the resolution in high density regions to ensure that the Jean's criteria was met, finding their method was robust and comparable to results from AMR calculations. Importantly they used 13 children for each parent ($n_{\rm child} = 13$ where we refer to more massive and less massive particles as `parents' and `children', respectively) to ensure spherical symmetry in 3D when splitting particles and established an empirical test to determine the radius of the sphere that the children are located on ($r_{\rm sep}$). \citet{Kitsionas:2007ih} showed that using a mass weighted kernel (with 50 times the mass of the largest particle) was preferable to volume weighted. 

\citet{FeldmanBonet:2007bh} introduced a general splitting procedure, placing child particles symmetrically in hexagons or triangles around their parent to conserve angular momentum. Importantly, they quantified the error that is generated when a parent particle is split into children; when the children have equal mass this error is due to the new particle arrangement and is a function of the distance between the children and their parent. 

\citet{Lastiwka:2005bh} showed that this error could be reduced by slightly shifting the children. \citet{Lopez:2013nj} extended this idea by measuring the error between the parent and child distributions as a function of the separation between the children when they are newly placed, $r_{\rm sep}$. Solutions to this error include shuffling the particles until the error is minimised \citep{Vacondio:2013wd}, blending the region between the parent and children regions \citep{Barcarolo:2014vu}, particle disordering \citep{Chiron:2018hq} and particle regularisation \citep{Gao:2022wd}. \citet{Vacondio:2013wd} first introduced particle merging in incompressible applications but found it too expensive for practical use. \citet{Chiaki:2016qs} showed that the noise introduced by a split could be reduced by placing the children particles on the vertices of a Voronoi mesh. 

Most recently, \citet{Franchini:2022nh} used particle splitting in a meshless-finite-volume method \citep{Hopkins:2014qy} to better resolve circumstellar discs within the inner cavity of a circumbinary disc. Adopting the method from \citet{Angles-Alcazar:2021jj} they used $n_{\rm child}=2$ and $r_{\rm sep} = \rm{min}(0.25 \times \textit{h}_{\rm parent}, 0.35 \times \textit{d}_{\rm neighbour})$ (where $d_{\rm neighbour}$ is the distance to the nearest neighbour) to ensure that split particles were not inadvertently placed too close to existing particles (i.e. that the fluid elements did not overlap). The robustness of their method was demonstrated in \citet{Duffel:2024ng} where their splitting method found agreement with comparable codes including \textsc{Phantom} \citep{Phantom}.

While the above implementations are promising, practical limitations remain. First, while most authors state that merging is possible, in practice it is not used --- and never in simulations of compressible flow. On the rare occasion that merging \emph{is} included it is found to be computationally prohibitive \citep{Vacondio:2013wd} which restricts what problems APR can be practically applied to. Second, typical implementations require hard-wiring of parameters like $r_{\rm sep}$ in order to reduce noise when particles are split, but these values are  derived from simple tests \citep[e.g.][]{Kitsionas:2002bh}. Third, some of the above applications in astrophysics allow particles of different refinement levels to mix which may result in numerical instabilities \citep[e.g.][]{Chiron:2018hq}.

In this paper we introduce an APR implementation into the smoothed particle hydrodynamics code \textsc{Phantom} that includes both splitting and merging, separation of refinement levels and which is accurate and fast.
In Section~\ref{section:method} we outline the method. Section~\ref{section:examples} applies our method to examples and practical calculations compared to equivalent global high resolution simulations. We measure speed, accuracy and disc storage. In Sections~\ref{section:discussion}~and~\ref{section:conc} we discuss and draw  conclusions. Basic tests for the interested reader are summarised in \ref{section:boxtests}.

\begin{figure*}[ht]
    \centering
    \includegraphics[width=0.8\textwidth]{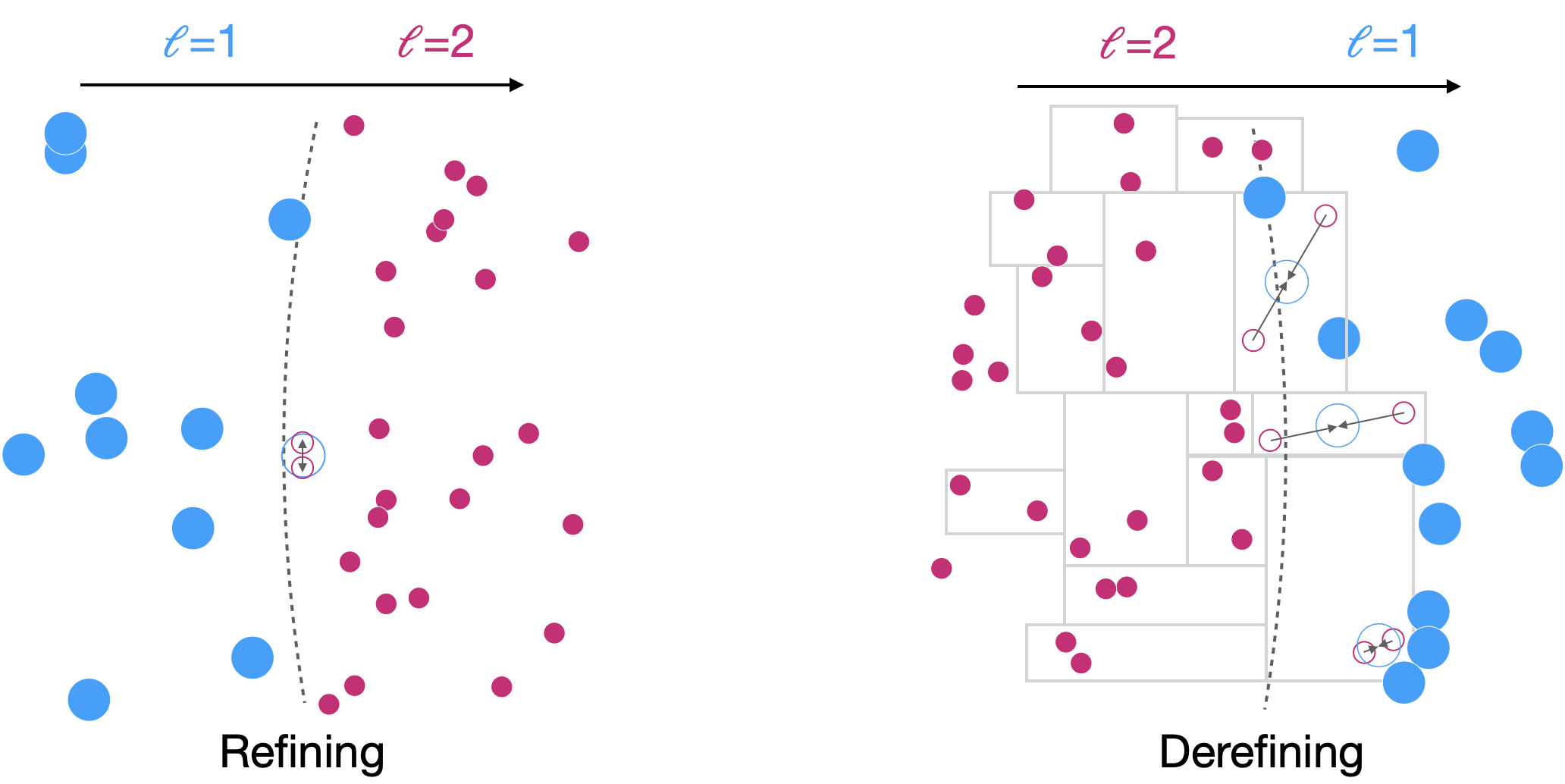}
    \caption{Schematic showing our refining and derefining process. The flow of the fluid is left to right and parent particles are larger and blue, children particles smaller and pink and the particle size is proportional to its mass. Particles that are split or merged in the time step shown are indicated with outlines and $\ell$ shows the refinement level. \emph{Left:} As a parent particle crosses the boundary it is split into two children particles, their common centre of mass is at the parent's location and they are split tangentially to the boundary. \emph{Right:} Children particles are paired according to our modified $k$-d tree grouping (indicated with the grey boxes). When the centre of mass of a cell crosses the boundary the particles are merged, with the parent adopting the average velocity and position of the children. Further details are in Section~\ref{section:method}. 
    }
    \label{fig:apr_schematic}
\end{figure*}

\section{Method}
\label{section:method}
Here we describe the core of our implementation including splitting, merging, relaxing and the order in which these are completed. The splitting and merging processes are summarised in Figure~\ref{fig:apr_schematic}.

\subsection{Overall procedure}
We assume spherical refinement and de-refinement zones. Their size and location may be either fixed or dynamic and co-moving with another particle. In some of our later applications the refinement zone is centred on a moving point mass particle but it may also be set by a particle property like density. Here we show examples with spherical zones but other volumes are possible. The main points of our implementation can be captured with five `rules':
\begin{enumerate}
    \item We assign all particles a refinement level, $\ell$, determined exclusively from the spatial position of the particle. The refinement level represents the number of refinements above the base resolution.
    \item When a particle enters a new refinement zone with a given refinement level
    \begin{itemize}
        \item it is split if the particle's current refinement level is less than that of the zone.
        \item it is merged if the particle's current refinement level is greater than that of the zone.
    \end{itemize}
    \item In simulations where the goal is to locally increase the resolution the default refinement level is set to $\ell = 0$. In simulations where we want to locally decrease the resolution the default refinement level is instead set to $\ell = \ell_{\rm max}$.
    \item We restrict the difference in mass between adjacent refinement levels to be strictly a factor of two.
    \item As \textsc{Phantom} stores the mass by particle type, the refinement level is also used to relate the refined mass $m_p(\ell)$ from the largest particle mass $m_p(0)$ with the following relation:
    \begin{equation}
      m_p(\ell) = \frac{m_p(0)}{2^{\ell}},
    \end{equation}
\end{enumerate}
where $\ell >= 1$. When we compare particles with different refinement levels we refer to the more highly resolved particles as `children' and the lower resolution particles as `parents'. In our scheme then, two children merge to form one parent and on a split one parent produces two children.

\begin{figure}
    \centering
    \includegraphics[width=0.8\textwidth]{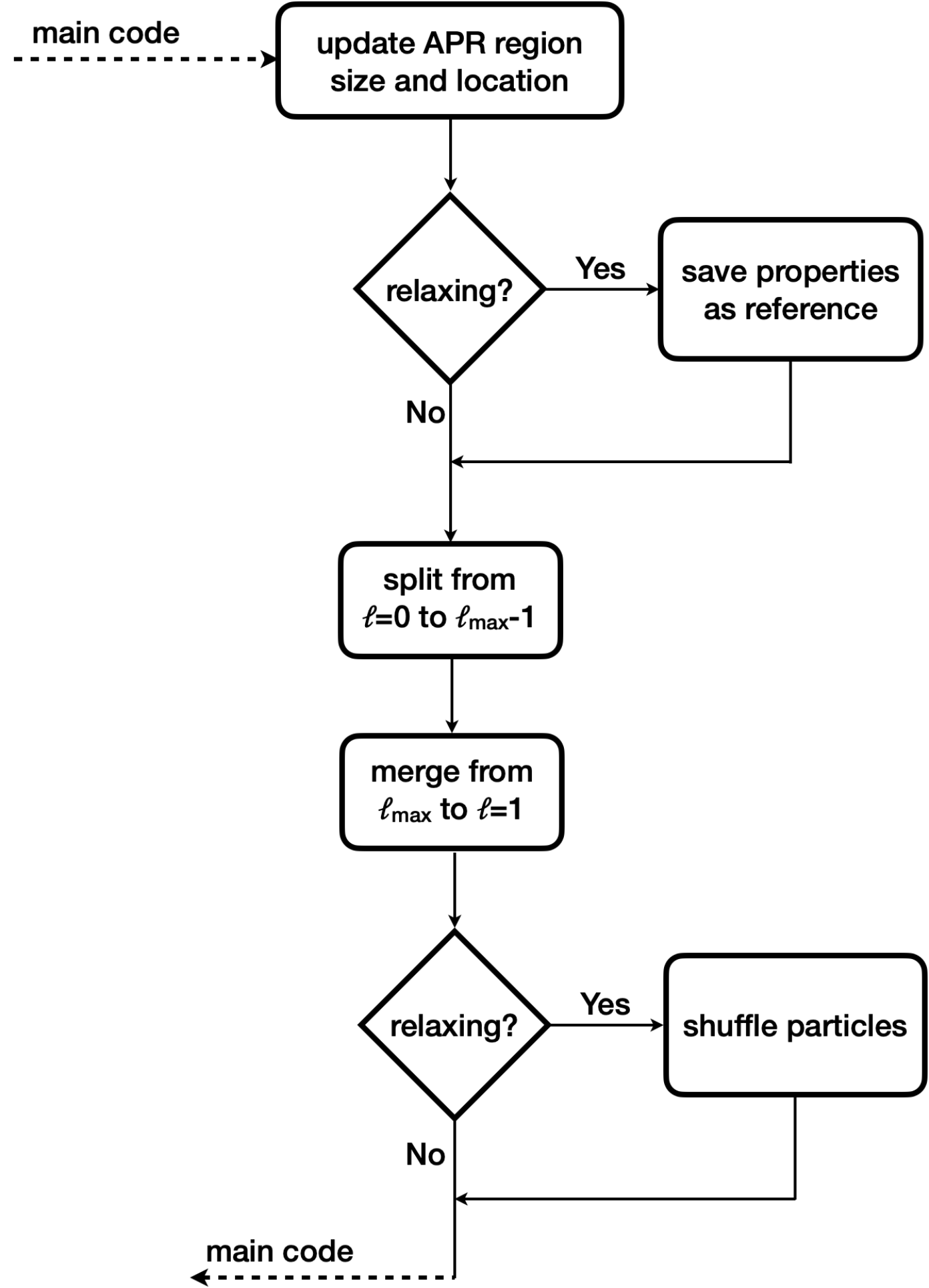}
    \caption{Flowchart summarising the APR routine implemented in \textsc{Phantom}.}
    \label{fig:apr_flowchart}
\end{figure}

\subsection{Splitting}
\label{section:split}
A parent particle with refinement level $\ell$ is split into two children when it crosses the boundary into a refinement zone with a higher refinement level. Once a particle is confirmed to be split;
\begin{enumerate}
    \item A new child particle is made with identical properties to the parent.
    \item The parent itself is reassigned to be a second child particle.
    \item Both children have their refinement level increased and their smoothing lengths rescaled (assuming constant density)
    \begin{equation}
        h_{\ell+1} = \left(\frac{1}{2}\right)^{1/3} h_\ell.
    \end{equation}
    \item Children are placed on opposite sides of the parent at a distance $r_{\rm sep} = 0.2 h_\ell$.
    \item Children are additionally placed tangential to the boundary of the refinement region (as in Figure~\ref{fig:apr_schematic}).
\end{enumerate}

The scaling value in $r_{\rm sep}$ has been determined empirically to reduce noise following the test from \citet{Kitsionas:2002bh}, however our inclusion of relaxing (see Section~\ref{section:relax}) means that our implementation is not particularly sensitive to this choice (as demonstrated in \ref{section:adjusting_split}). Additionally, as \textsc{Phantom} calculates the density and smoothing lengths iteratively \citep[2.1.4,][]{Phantom}, the smoothing lengths of surrounding particles are accurately adjusted to accommodate the new particles straight after the split (and later, merge) has occurred.\\

To ensure particles are split tangentially to the boundary in 3D we identify the vector $\vec{v}$ between the original particle and the centre of the refinement zone. A vector $\vec{w}$ represents the perpendicular bisector of $\vec{v}$ and is always tangential to the boundary of the refinement zone. We rotate $\vec{w}$ around $\vec{v}$ through a random angle and place the children on opposite sides of the parent along $\pm \vec{w}$. As in \citet{Franchini:2022nh}, if the distance shifted is more than $0.35\times$ the distance to the closest neighbour we adopt this value instead, however we have found that this only affects a handful of particles even in very well resolved simulations.

\subsection{Merging}
\label{section:merge}

We make use of a modified $k$-d tree in \textsc{Phantom} to make rapid on-the-fly merging possible. We have altered the existing $k$-d tree so that it always returns leaf nodes with exactly two particles (this modification is only used in the call from the APR routine and is not used elsewhere in the calculations). Every time the domain is split through the centre of mass we enforce an even number of particles on either side of the new “subdomains”. When a split occurs that results in an odd number of particles in each new subdomain, we move the location of the split across one particle in the direction that makes the most balanced split by number of particles. The tree continues until finally all leaves have precisely two particles.

At each time-step where merging occurs;

\begin{enumerate}
    \item All particles at the same refinement level $\ell$ are isolated and then paired them using our modified $k$-d tree.
    \item For each leaf in this tree, we calculate what refinement level the leaf has based on it's centre of mass and compare it to the refinement level of the child particles, $\ell$.
    \item The particles are merged when the refinement level based on the position of the leaf is less than the refinement level of the child particles in the leaf.
    \item One of these children particle is removed from the simulation and the remaining child particle is reassigned as a parent.
    \item The parent is given the position and velocity of the centre of mass of the leaf (i.e. the two children), the refinement level is decreased and the smoothing length is scaled by
    \begin{equation}
      h_{\ell-1} = 2^{1/3} h_{\ell}.
    \end{equation}
    \item If an odd number of particles are considered for merging, we discard the last particle that is listed at that timestep. In the next timestep when particles are again considered for merging this particle is then included.
\end{enumerate}

The right panel of Figure~\ref{fig:apr_flowchart} schematically shows the merging process. With our implementation it is important to note that although the boundary for where merging occurs is well defined, because we are merging based on the children particle distribution the practical boundary is fuzzy and a transition region typically about 10\% of the larger smoothing length naturally forms. In well resolved simulations (e.g. Section~\ref{section:examples}) this region is negligible in width. A major benefit of our approach is we have at most a maximum of one spare particle per refinement level per timestep. As we do not consider $\ell=0$ particles for merging, the maximum error $e$ in the total mass of these spare particles for $\ell_{\rm max}$ refinement levels is
\begin{align}
    e = m_p(0) \sum_{\ell=1}^{\ell_{\rm max}} \frac{1}{2^{\ell}},
    \label{equation:mass_error}
\end{align}
which is conveniently always less than $m_p(0)$, the largest particle mass.

\begin{table*}[t]
    \centering
    \begin{tabular}{ccccccc}
        Name & $r_{\ell}$ (au) & $dr_{\ell}$ (au) & $N$ ($\times 10^6$) & Error (\%) & Speed up ($\times$) & Storage\\
         B1 & 15.00 & 20.0, 25.0 & 1.19 & $9.0\times10^{-3}$ & 2.32 & 0.15\\
         B2 & 20.0 & 30.0, 40.0 & 1.14 & $3.8\times10^{-3}$ & 1.96 & 0.14\\
         B3 & 40.0 & 50.0, 60.0 & 1.44 & $3.6\times10^{-3}$ & 1.80 & 0.18\\
         B4 & 80.0 & 90.0, 100.0 & 1.75 & $1.3\times10^{-3}$ & 1.07 & 0.22\\
         P1 & 0.50 & 0.75, 1.00 & 2.16 & 1.4 & 1.81 & 0.23\\
         P2 & 0.50 & 1.00, 1.50 & 2.09 & 4.5 & 1.76 & 0.23\\
         P3 & 1.00 & 1.50, 2.00 & 2.09 & 6.1 & 1.90 & 0.23\\
         P4 & 1.00 & 2.00, 3.00 & 2.20 & 7.8 & 1.53 & 0.25\\
         P5 & 2.00 & 2.50, 3.00 & 2.11 & 9.1 & 1.74 & 0.24\\
         P6 & 2.00 & 3.00, 4.00 & 2.32 & 8.6 & 1.38 & 0.27\\
         F1 & 40.0 & 45.0, 50.0 & 0.61 & $1.3 \times 10^{-4}$ & 6.62 & 0.15\\
         F2 & 40.0 & 50.0, 60.0 & 0.62 & $2.0 \times 10^{-4}$ & 6.14 & 0.15\\
         F3 & 60.0 & 70.0, 80.0 & 0.63 & $1.5 \times 10^{-4}$ & 5.47 & 0.16\\
         F4 & 80.0 & 90.0, 100.0 & 0.64 & $1.5 \times 10^{-4}$ & 5.28 & 0.16\\
         F5 & 60.0 & 70.0, 80.0 & 5.7 & - & - & -\\
         F6 & 60.0 & 50.0, 60.0, 70.0, 80.0 & 3.0 & - & - & -\\
    \end{tabular}
    \caption{Summary of the APR simulations shown in Section~\ref{section:examples}. Columns state the Name of the simulation, the radius of the central refinement region $r_{\ell}$, the steps into each refinement zone $dr_{\ell}$ and the average number of particles used $N$. The error, speed up and storage are all compared to the high resolution reference cases. The method to measure the error is described in the text for each simulation and is measured according to Equation~\ref{equation:error}. The storage is calculated as a fraction compared to the high resolution reference calculation. All simulations have 3 levels of increased refinement except F6 which has $\ell=6$. Here `B' refers to binary, `P' to planet-disc and `F' to flyby simulations.}
    \label{tab:simulation_summary}
\end{table*}

\subsection{Relaxing}
\label{section:relax}
When splitting or merging a significant number of particles – for example the first time the splitting/merging routine is called – there is a discrepancy between the original and refined density distribution. This effect was identified by \citet{FeldmanBonet:2007bh} and is due to how well the new particle arrangement can represent the original density distribution.

To combat this we have implemented a relaxing procedure which shuffles the introduced particles until they more accurately represent the original distribution. When splitting or merging an original set of particles into a new set, our algorithm 
\begin{enumerate}
\item Calculates the accelerations of the new set of particles at their current locations, $a_{\rm new}$.
\item Calculates the accelerations at the locations of the new set of particles, interpolated from the original reference particles, $a_{\rm ref}$.
\item Shifts each particle by $\Delta x$ calculated from
\begin{align}
    \Delta x = 0.5 \Delta t^2 (a_{\rm new} - a_{\rm ref}),
\end{align}
where $\Delta t$ is calculated from the sound speed and smoothing length by $\Delta t = 0.3 h/c_s$ on the new particle.
\end{enumerate}

For any given particle the magnitude of the shuffle is also capped to be less than the particles smoothing length. At each shuffle step, we estimate a `kinetic energy' defined as the magnitude of the shift divided by the particle’s timestep squared and summed across all the shuffled particles. For multiple shuffles we find that the kinetic energy decreases in an exponential fashion with each subsequent shuffle smaller than the previous. Our shuffling process is repeated until either the total kinetic energy for a shuffle has decreased to 0.5\% of its original value or 50 shuffles have been completed. These limits are set to strike a balance between accuracy and computational expense, as we have found that more shuffles does not continue to dramatically improve the particle distribution. How quickly the kinetic energy limit is reached does depend on the resolution, with higher resolution simulations achieving it in fewer shuffles.

\subsection{With individual timesteps}
A key part of our implementation is compatibility with individual particle timesteps as these are used widely in \textsc{Phantom} applications \citep[see][]{Phantom}. Each particle is assigned to a timestep bin; these are arranged such that the zeroth bin has $\Delta t$ = $\Delta t_{\rm max}$ (which is limited to be the time between outputs) and particles are arranged in bins according to their local timestep constraint, where $\Delta t$ decreases by factor of two in each bin.

Each bin is then evolved separately, synchronising with the bin above when the appropriate timesteps are synchronised.
Particles are then defined as `active' (as in, they will be moved by the stepping routine) when their bin is being evolved.

When implemented in conjunction with APR we restrict the splitting and merging procedure to occur over active particles only. This does mean that a merging particle's closest neighbour could be on a different time-step, forcing the particle to merge with an active particle further away. In practice this is rare because the particles have been paired using our modified $k$-d tree. This pairing takes advantage of the fact that spatially associated particles tend to have the same acceleration and so are naturally on the same timestep bin such that the closest neighbour is also on the same timestep. Additionally, when a particle is either split, merged or a new particle introduced through either of those procedures it is conservatively assigned the shortest timestep (corresponding to the maximum timestep bin).

\subsection{Every timestep}
Figure~\ref{fig:apr_flowchart} summarises the process undertaken at each timestep on active particles. First the location, size and number of refinement regions are updated. If relaxing is required at this step the properties of all particles are also saved as a reference.

We then consider whether any particles should be split. Using the particles current position we calculate what refinement level it should have and compare this to its current refinement level. When the particle has a refinement level less than what it should have we split it according to Section~\ref{section:split}. We repeat this process across for all active particles and for each refinement level, starting from the lowest refinement level.

After all the particles have been split we then consider whether any should be merged. We isolate all particles that are at a particular refinement level and put these particles into our modified tree routine, merging as per Section~\ref{section:merge}. Mirroring the splitting procedure, this process is repeated from the highest to lowest refinement level. This order means that particles are only ever changed one refinement level each split/merge which we have found leads to a smoother distribution (e.g. \ref{section:four_is_too_many}). Only once all the particles have been split and merged do we consider the relaxing process, using the original particle distribution as the reference set. With the particles refinement levels updated the program then returns to the normal time stepping routine and continues.

\begin{figure*}[ht]
    \centering
    \includegraphics[width=\textwidth]{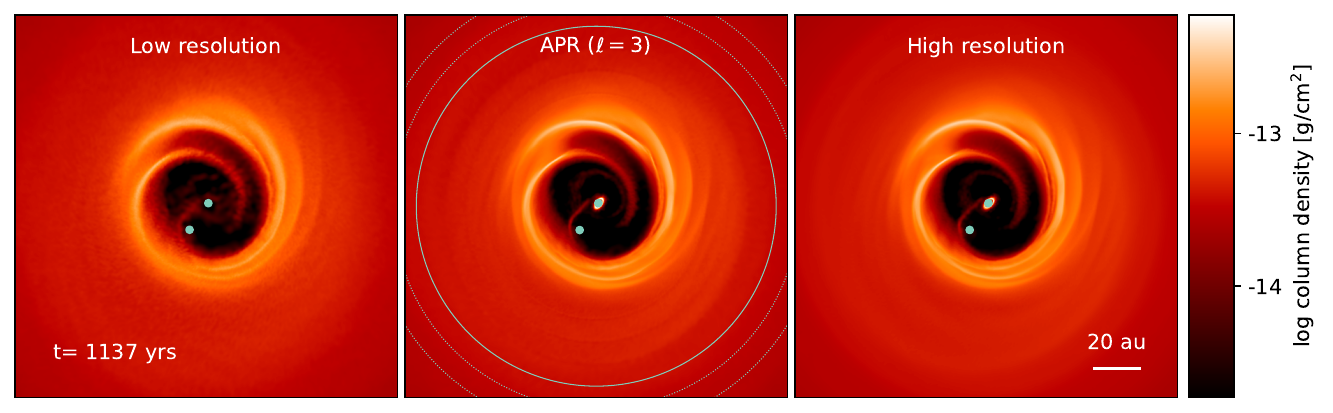}
    \caption{Column density of the HD142527 simulations from \citet{Price:2018pf} at $t=1137$ years. The left and right panels have no APR and are separated by a factor of two in spatial resolution. The middle panel shows simulation B4 with three levels of refinement, locally matching the evolution of the high resolution reference case. The refinement zone is shown with the light green circles where the highest resolution region is indicated with a solid line and the nested regions with dotted lines. The similarities in the streamers and circumprimary disc confirm our APR implementation is capable of accurately locally increasing the resolution of a simulation. A movie of these simulations is available \href{https://youtu.be/5gSmvOKuiLA}{online}.}
    \label{fig:HD142527_rendered}
\end{figure*}

\begin{figure}
    \centering
    \includegraphics[width=\textwidth]{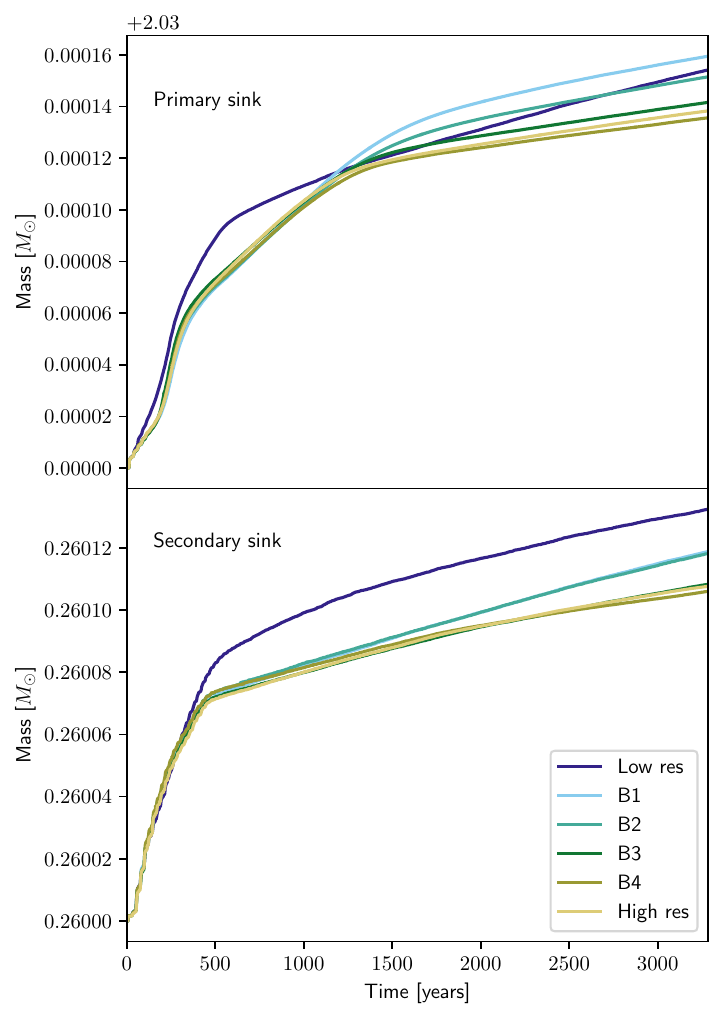}
    \caption{Mass of each sink in the HD142527 simulations with different combinations of refinement region sizes. The mass accretion rate reflects the properties of the disc surrounding the sink, confirming the structure of circumprimary disc in the APR simulations is the same as the high resolution reference calculation (as seen in Figure~\ref{fig:HD142527_rendered}).}
    \label{fig:HD142527_mass}
\end{figure}

\section{Example applications}
\label{section:examples}
We demonstrate the accuracy, speed up and typical use cases of our APR implementation with three example applications: a circumbinary disc, planet disc interactions and a stellar flyby. In each case we include a low and high resolution comparison simulation, where the high resolution simulation has the same global resolution as the highest APR zone. Our examples also demonstrate the adaptability of the implementation and provide typical values for the size of the refinement region and the step width. {Our examples include gravity from sink particles (stars and planets) as well as accretion onto those sinks. We also assume a vertically isothermal temperature profile for these applications, but we refer to \ref{section:boxtests} for tests with an adiabatic equation of state. Radiative cooling or source terms that are sensitive to temperature are left to future work. Our results are summarised in Table~\ref{tab:simulation_summary}. In all cases we find that our implementation is accurate, fast and requires less storage space when compared to the high resolution reference cases.

In the examples shown here we choose to use $n_{\rm child} = 2$ and nested refinement levels and with `relaxing' only implemented in the first step (see \ref{section:boxtests}). We additionally ensure that the velocity of the fluid across the boundary is small (see \ref{section:aprzone_size}) by choosing the location of our refinement regions carefully. These examples are presented with a Wendland C2 kernel with a kernel radius of $2h$ and $h_{\rm fact} = 1.3$ \citep{Wendland:1995hd,Phantom}. As we are considering particles of different masses co-existing in the same simulation it is prudent to consider kernels other than the typical $M_4$ cubic spline \citep[e.g.][]{Denhen:2012up}. Our comparison of $M_4$, Wendland C2, Wendland C4 and Wendland C6 is shown in \ref{section:boxtests}. Here we find that the $M_4$ cubic spline is robust to pairing in these tests but that the higher order splines offer a slightly smoother transition between the refinement zones. We thus chose the Wendland C2 as a balance between computational cost and accuracy but note that the cubic spline is likely sufficient for most purposes.

\begin{figure*}[t]
    \centering
    \includegraphics[width=\textwidth]{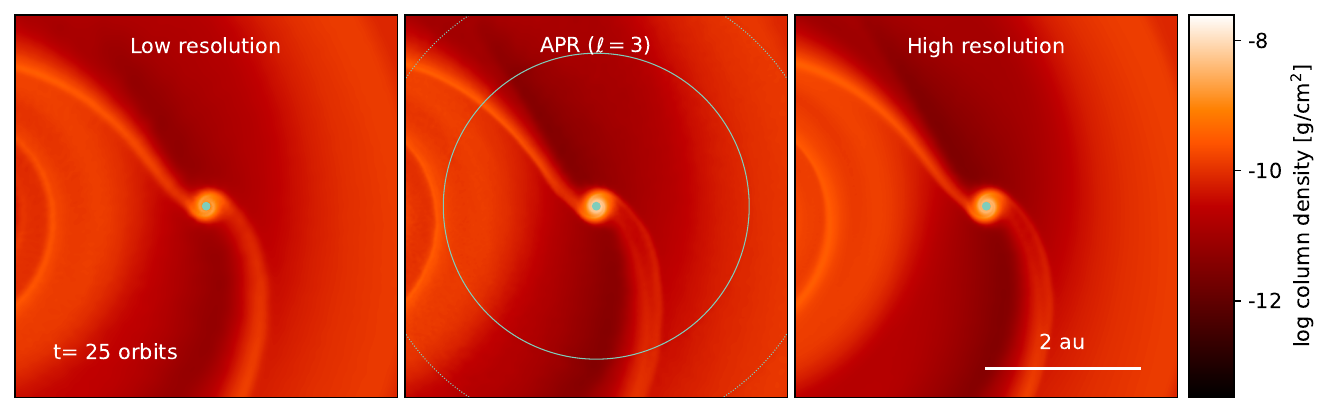}
    \caption{Column density of the region surrounding a planet embedded in a disc shown in the corotating frame. The left and right panels have no APR and are separated by a factor of two in spatial resolution. The middle panel shows simulation P6 with three levels of refinement, matching the circumprimary disc structure and spiral arms in the high resolution reference simulation. The refinement zone is shown as in Figure~\ref{fig:HD142527_rendered}. This test demonstrates that structures like spiral arms are faithfully reproduced even when they cross the refinement boundaries. A movie of these simulations is available \href{https://youtu.be/ggBkwEpk2Q8}{online}.}
    \label{fig:planets_rendered}
\end{figure*}

\begin{figure}[h!]
    \centering
    \includegraphics[width=\textwidth]{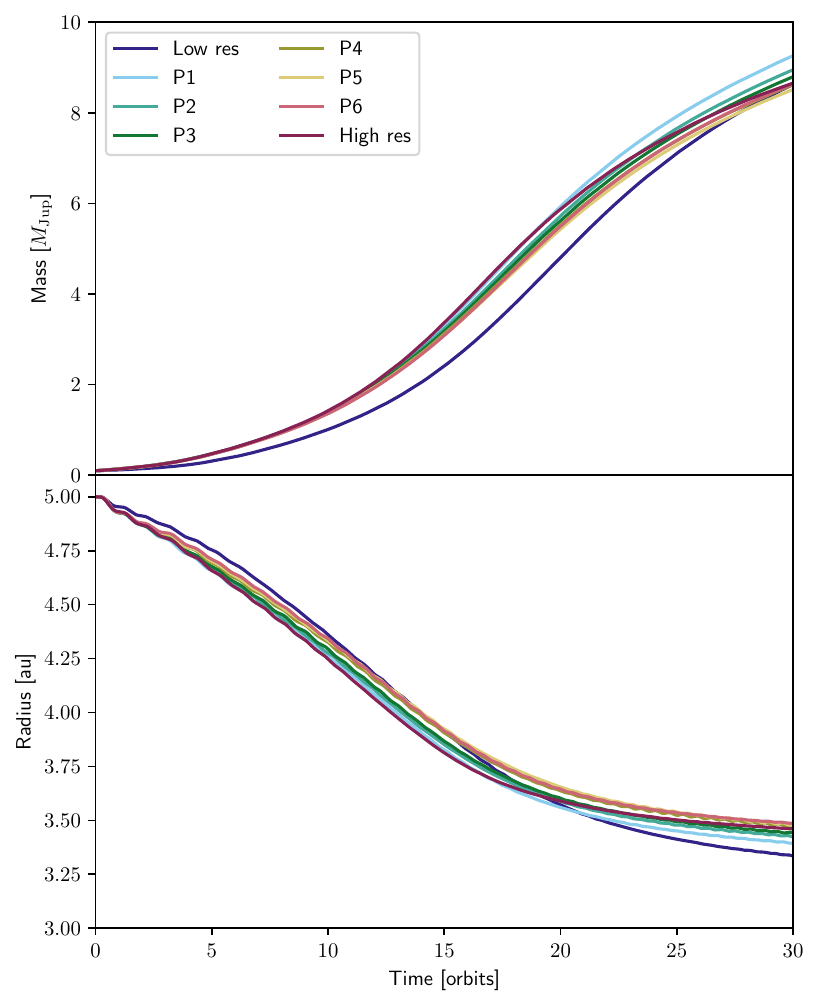}
    \caption{Mass and radius of the planet in the planet-disc simulation with different combinations of refinement region sizes. The APR simulations and high resolution simulations are again distinct from the low resolution reference simulation.}
    \label{fig:planets_mass}
\end{figure}

\subsection{Circumbinary disc}
We repeat the concluding simulation from \citet{Price:2018pf} of a circumbinary disc nominally representing the protoplanetary disc HD~142527 \citep[but see][]{Nowak:2024gh}. The star and disc parameters are the same as used in these works and our low resolution reference simulation has $N=1\times 10^6$ particles, as in their work. Our high resolution reference simulation is the same as the low except it has double the spatial resolution with $N=8\times 10^6$ particles. In our APR versions (simulations B1-B4) the refinement region has three nested levels of refinement ($\ell=3$) and is centred on the centre of mass of the two stars so that the accretion streams and circumstellar discs are more highly resolved. As the gas has a negative radial velocity, we find that this application mostly represents a test of splitting --- little if any merging occurs in these simulations. This is similar to the published applications of APR in GIZMO \citep{Franchini:2022nh,Duffel:2024ng}.

Figure~\ref{fig:HD142527_rendered} shows a representative APR simulation of HD~142527 book-ended by reference calculations. At $t=1137$ years in the low resolution reference calculation we find a poorly resolved circumprimary disc consistent with \citet{Price:2018pf}, but this disc is recovered in the high resolution case on the far right. Our APR simulation shows the same disc structure around the primary star as in the high resolution case. The spiral arms at the inner edge of the disc are well recovered when they are within the highest refinement region (e.g. simulations B3 and B4). The consistency between the APR and high resolution case is maintained through the end of the simulation at $t=3283$ years.

Figure~\ref{fig:HD142527_mass} shows how the mass of the two sink particles changes throughout the reference simulations and simulations B1--B4. The evolution of the mass for the APR simulations closely follows the high resolution reference cases where the refinement region covers the cavity, indicating that the disc structure and thus the accretion rate is the same across these simulations. The low resolution simulation with the poorly resolved circumprimary disc accretes faster, resulting in a larger final mass. Simulations B1 and B2 show distinct behaviour after $t\sim1000$ years with up to $\sim3$ times more mass accreted onto both sinks due to their small refinement regions; the spiral arms and even the secondary sink are not consistently resolved inside the highest refinement region. For both B1 and B2 the lower resolution of the inner edge of the cavity then leads to about $30\%$ more mass  falling onto the sinks and thus the higher accretion rate for both.

We calculate the error, $\mathcal{E}$, using the largest percentage difference in mass between the APR versions and high resolution reference case on each sink as
\begin{align}
    \mathcal{E} = {\rm max} \left[\frac{m_{\rm APR}(t) - m_{\rm ref}(t)}{m_{\rm ref}(t)} \times 100\right].
    \label{equation:error}
\end{align}
We repeat this calculation across both sinks and take the largest of these as the error in our simulation when compared to the high resolution reference case, summarised in Table~\ref{tab:simulation_summary}. We find the largest error of $9.0\times10^{-3}$\% lies with the secondary sink of the B1 simulation, possibly because it ventures closer to the edge of the refinement region each orbit. In contrast to our other example applications, our APR simulations here show a speed-up compared to the globally high resolution case of $\sim 1.1-2.3\times$ faster (Table~\ref{tab:simulation_summary}). The speed up is inversely proportional to the size of the refinement zone because it is centred on the binary; the tightest and most computationally expensive orbits are located here and our APR simulations add more particles to this region.

\subsection{Planet-disc interaction}
We simulate the interaction of a planet embedded in a disc where the region around the planet is refined using APR. The disc extends from $R_{\rm in}=1$ au to $R_{\rm out}=10$ au with a total disc mass of $0.05$M$_{\odot}$. We adopt typical protoplanetary disc values with a surface density profile $\Sigma \propto (1 - \sqrt{R_{\rm in}/R}) R^{-p}$ with $p=1$, a sound speed $c_{\rm s} \propto R^{-q}$ with $q=0.25$, a disc aspect ratio of $H/R = 0.05$ at the inner edge $R=R_{\rm in}$ where $H\equiv c_{\rm s}(R)/\Omega(R)$ is the scale-height, $\Omega(R) = \sqrt{GM_*/R^3}$ is the Keplerian angular velocity, $M_* = 1$M$_\odot$ is the central mass and $R$ is the cylindrical radius. We also assume a \citet{shakura_sunyaev} viscosity of $\alpha=0.005$, modelled with the method outlined in \citet{lodato_2010}. Into this disc we add a planet with $0.1$M$_{\rm J}$ initially located at $R_{\rm p}=5$ au, which is large enough to generate spirals in the disc but not so massive that the wakes of the spirals interact strongly with the planet. We set the accretion radius of the sink to be $0.25 R_{H} = 0.04$ au where $R_H$ is the Hill radius \citep{Nealon:2018ic}. Our low and high reference simulations use $N=1 \times 10^6$ and $N=8\times 10^6$ respectively. We conduct six APR simulations (P1-P6, see Table~\ref{tab:simulation_summary}) varying the size of the refinement region and the size of the step around it, all with a maximum refinement level of $\ell=3$ (corresponding to the high resolution reference).

Our simulations with a planet show the formation of spiral arms, indications of a circumplanetary disc and a shallow gap being carved. Figure~\ref{fig:planets_rendered} compares the column density of our low and high resolution reference case to the representative example P6.  For all of these simulations we find that the spiral arms generated by the planet are recovered faithfully even as they cross multiple refinement levels. The column density contrast of the circumplanetary disc and the spiral arms are visually almost identical between our APR simulations and the high resolution reference case.

Figure~\ref{fig:planets_mass} shows the mass of the planet and radius measured from the central star. As with the HD~142527 simulations, the high resolution reference and the APR versions are qualitatively distinct from the low resolution case. Using Equation~\ref{equation:error} we calculate the error from the mass but also equivalently from the radius, taking the largest percentage difference as the error reported in Table~\ref{tab:simulation_summary}. We find the largest difference of 9.1\% in the mass and 3.06\% in the radius, both for simulation P5. While this error is the largest we recover with our APR implementation in the examples shown here it is still about $1/3$ of the largest difference between the low and high resolution reference cases. Additionally, \citet{Nealon:2018ic} showed that the choice of accretion radius has a stronger impact on the planet location and mass (their Figure A1) than the difference we have found here. In general we find that the error increases with both the size of and the width of the refinement region.

\begin{figure*}[ht]
    \centering
    \includegraphics[width=\textwidth]{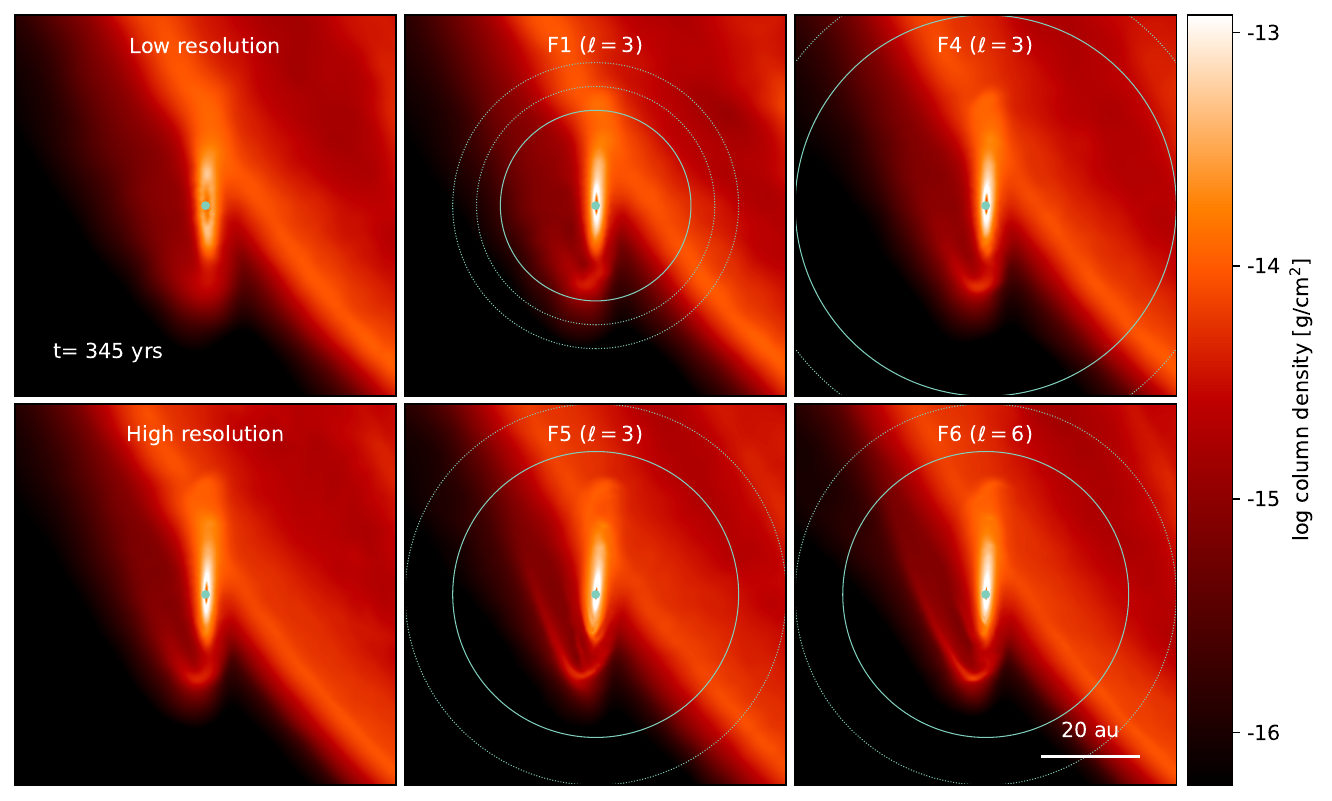}
    \caption{Comparison of discs formed from captured material around the perturber in the flyby simulation \citep[see][]{Smallwood:2024yq}. The `Low resolution' and `High resolution' panels do not have APR and are separated by a factor of two in resolution. Simulations F1 and F4 initially have $N=5\times10^5$ particles with 3 levels of APR, F5 has $N=4\times10^6$ particles with 3 levels of refinement and F6 has $N=5\times10^5$ particles with 6 levels of refinement. The refinement zone is shown as in Figure~\ref{fig:HD142527_rendered}. The disc structure is similar irrespective of the base resolution of the simulation but the tidal stream onto the disc depends on the size of the refinement region and the number of levels. A movie of these simulations is available \href{https://youtu.be/8Uj-wv5TTX8}{online}.}
    \label{fig:flyby_panel_rendered}
\end{figure*}

\begin{figure}[h!]
    \vspace{0.1cm}
    \centering
    \includegraphics[width=\textwidth]{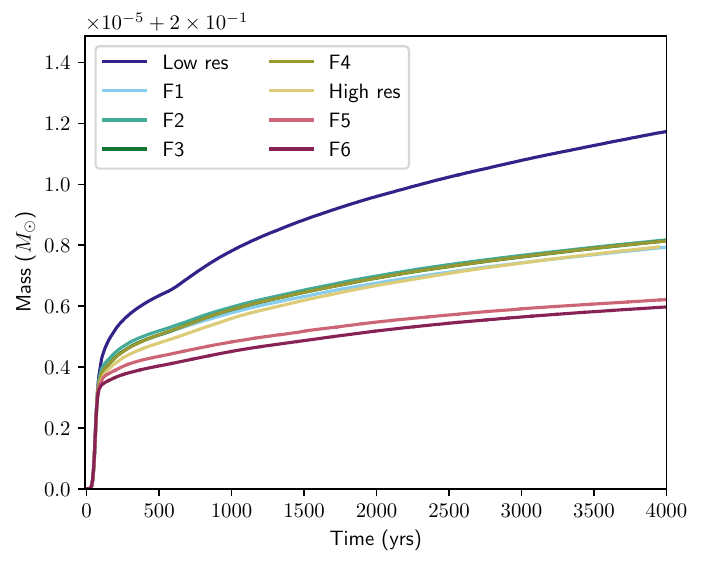}
    \caption{Mass of the perturber star in the flyby simulation for different combinations of refinement regions and maximum refinement levels. As before, increasing the resolution using APR leads to a lower mass accretion rate and the similarities between the APR and high resolution reference evidence the similarity in the disc structure around the perturber star.}
    \label{fig:flyby_mass}
\end{figure}

\begin{figure}[h!]
    \vspace{0.1cm}
    \centering
    \includegraphics[width=\textwidth]{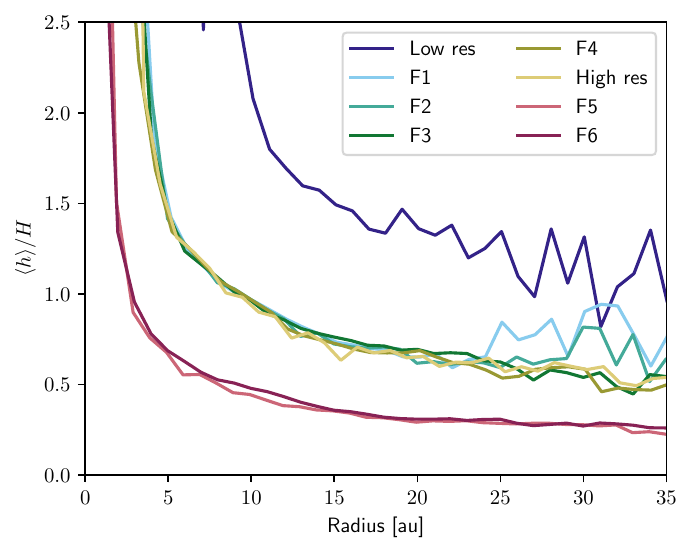}
    \caption{Quantifying the resolution of the perturber disc in the flyby calculations using $\langle h \rangle /H$ as a function of $R$. Three levels of refinement corresponds to a factor of two in linear resolution and this is recovered here by the $\langle h \rangle /H$ decreasing by about half for three refinement levels. Colour scheme is the same as in Figure~\ref{fig:flyby_mass}.}
    \label{fig:flybys_honH_and_tilt}
\end{figure}

\subsection{Flyby encounter}
\citet{Smallwood:2024yq} showed that there is a direct and robust relationship between the orientation of the disc formed from captured material in a flyby encounter and the original disc. With a broad suite of simulations, their results showed that the disc that formed around the perturber was always tilted twice as much as the original inclination between the perturber and the disc around the primary star. However the perturber disc in their simulations was generally underesolved (see their Figure A2). \citet{Smallwood:2024yq} conducted a resolution study with $5\times10^5$ and $4\times10^6$ particles and determined that in order to well resolve the disc around the perturber they would need to model the original disc with $\sim 144 \times 10^6$ particles.

We repeat the calculation from \citet{Smallwood:2024yq} at $5\times10^5$ particles for our low resolution reference and $4\times10^6$ as our high resolution reference. We then conduct four APR variations (F1 - F4) with the refinement region centred on the perturber star. The upper panel of Figure~\ref{fig:flyby_panel_rendered} shows the column density of the two reference cases and simulations F1 with a refinement region radius of $20$ au and F4 with $40$ au, about $300$ years after pericentre passage \citep[as in][]{Smallwood:2024yq}. The two APR cases and the high resolution share the same disc structure but the tidal stream is different. In F1 with the smallest refinement region the tidal stream more closely resembles that of the low resolution reference simulation, in line with our expectations.

As before, Figure~\ref{fig:flyby_mass} shows the mass of the perturber sink throughout the simulations. We find that the high resolution reference case and APR versions describe the same mass accretion pathway and that it is distinct from the low resolution reference case. The error measured from the mass of the perturber sink is $\lesssim 10^{-3}$\% for all combinations of APR used here and with speed ups of between $5.28-6.62\times$ faster than the high resolution reference case.

As in \citet{Smallwood:2024yq} we also measure the resolution of the perturber disc for the different simulations. Figure~\ref{fig:flybys_honH_and_tilt} shows the $\langle h \rangle/H$ resolution at $t=2500$ years after the pericentre encounter where $\langle h \rangle/H$ is the shell averaged smoothing length. We find that the high resolution reference case and APR versions produce the same profile, confirming that they are the same local resolution. In particular, the profiles for F1 and F2 deviate around $R=25$au and $30$ au respectively, near to the edge of the refinement region for these simulations. The factor of $1/2$ in $\langle h \rangle /H$ between the APR simulations and the low resolution case demonstrates that the resolution has doubled (as expected from three refinement levels).

These example applications so far have demonstrated the accuracy and speed of the APR implementation by comparing to a globally high resolution reference case, but this is not the most powerful or intended use of APR. Here we perform an additional two simulations using APR on the flyby problem: simulation F5 which initially has $N=4 \times 10^6$ particles and uses three levels of refinement ($\ell=3$) and simulation F6 which initially has $N=5 \times 10^5$ particles and uses six levels of refinement ($\ell=6$). Around the perturber these simulations achieve the same local resolution in terms of local particle number.

The lower row of Figure~\ref{fig:flyby_panel_rendered} shows the discs formed around the perturber star for F5 and F6, finding the same disc structure as in the previous cases. These two simulations are also included in Figures~\ref{fig:flyby_mass} and \ref{fig:flybys_honH_and_tilt}. The mass accreted onto the sink in simulations F5 and F6 is $9.2\times 10^{-4}$\% lower 
than the high resolution reference case. This difference is about half that between the low and high resolution reference cases ($1.89 \times 10^{-3}$\%) and consistent with our previous observation that increasing the resolution decreases the mass accretion rate. Figure~\ref{fig:flybys_honH_and_tilt} demonstrates that the linear resolution of these two simulations is double the previous F1 - F4 simulations. We have thus shown that with just $5.7\times10^6$ (F5) and $3.0\times10^6$ particles we can achieve an equivalent resolution around the perturber of simulation that uses $32\times10^6$ particles globally.

\section{Discussion}
\label{section:discussion}
Our results demonstrate that our method is accurate and fast (see Table~\ref{tab:simulation_summary}) with the most rapid speed up seen in the flyby simulations and the slowest in the circumbinary disc simulations. In the flyby case the majority of the splitting/merging occurred during the actual flyby encounter and the time-step constraining orbits were around the primary star which we did not change in our APR simulations. Together, these led to a significant speed up in the simulation. By contrast, in the circumbinary disc we increased the resolution in the region of the simulation where particles had the largest acceleration and thus the shortest time-steps. We still found a speed up in that case because of our use of individual time-steps \citep{Phantom}. In the planet-disc interaction simulations we found constant splitting and merging occurred as the variation in the orbital speed meant particles were continually entering and exiting the refinement region. For those simulations the size of the refinement region generally dictated the speed up, with smaller regions corresponding to faster speed ups.

The circumbinary disc and flyby tests demonstrated the importance of \emph{where} to set the refinement region. In the former, the largest error was measured when the inner edge of the circumbinary disc was not adequately resolved (simulations B1 and B2) such that material was entering the cavity with the same rate as the low resolution reference case and occasionally, the secondary star popped out of the refinement region. But when the inner edge of the circumbinary disc and the complete orbit of the secondary was captured in the refinement region (B3 and B4) the error decreased significantly as compared to the high resolution reference case. This effect was also noted by \citet{Franchini:2022nh}, who recommend setting the inner edge of the refinement region to $4\times$ the semi-major axis.

Our implementation is also designed to be as adaptable as possible. Our use of the $k$-d tree in \textsc{Phantom} for merging means that our APR method can naturally be used for de-refinement, where the APR zone has a lower resolution than the global simulation.
This is particularly useful when increases in density result in very small time-steps that effectively kill a simulation - commonly experienced in simulations including self-gravity \citep[e.g.][]{Longarini:2023ju,Lau:2022sd,Hall:2017gw}. Our method is also easily extendable to multiple refinement regions (e.g. refining separately around two planets in a disc).

An important and desirable property of SPH is its conservation properties. Conservation of kinetic energy and both linear and angular momentum are respected with APR when the children particles inherit the velocity of the parent \citep{FeldmanBonet:2007bh,Lopez:2013nj}. While our inclusion of relaxing means that children may not necessarily be placed symmetrically around the parent particle this only affects the conservation of angular momentum. Our method perfectly conserves mass and in simulations with accretion (where we may get an odd number of particles to be merged) our implementation caps the number of unmerged particles to be $\ell_{\rm max} - 1$ (Equation~\ref{equation:mass_error}). In addition to this, APR in \textsc{Phantom} is subjected to the regular conservation checks \citep[see Section 2.2,][]{Phantom}.

The current limitation of our method is a $\sim$5\% blip in density that occurs every time particles are either split or merged. In our 3D simulations this `blip' in density is cosmetic (it can just be seen in the middle panel of Figure~\ref{fig:planets_rendered}) but is more obvious in our tests on small amplitude linear sound waves outlined in ~\ref{section:boxtests} (although it does not seem to affect the propagation of the wave itself). We tested different methods to mitigate this feature and found that reducing $n_{\rm child}$ was ultimately the most effective measure, setting our limitation of $n_{\rm child}=2$ and thus having nested refinement regions. Importantly, the `blip' feature is also visibly present in the GIZMO implementation of APR \citep[see density renderings in][]{Duffel:2024ng} which demonstrates that it is not a function of our method, but is inherent to the process of changing the particle mass.

The most promising method to remove this blip is the blending method outlined by \citet{Barcarolo:2014vu} which allows particles to be introduced in a way that removes the noise before they contribute to the density summation. Although successful for incompressible flows, blending is practically very difficult to implement with compressible flows due to mass conservation in the blending zone. For example, we could successfully incorporate blending as in \citet{Barcarolo:2014vu} in our initial conditions, but once particles started moving through the blending region we found that the solution deteriorated rapidly. We additionally implemented other forms of relaxing \citep[e.g.][]{Lind:2012bj,Diehl:2015vu,Sun:2017jo} and other forms of blending \citep{Chiron:2018hq,Gao:2022wd} but did not find a satisfactory improvement on the blip found in the simulations.

\section{Conclusion}
\label{section:conc}
We introduce a live adaptive particle refinement implementation into the smoothed particle hydrodynamics code \textsc{Phantom}. We have considered example applications of a circumbinary disc, planet-disc interaction and a flyby to demonstrate our method. With these examples we have shown that our implementation is
\begin{enumerate}
    \item \emph{Accurate:} We measure the mass accreted onto sink particles as a proxy of the disc structure in our simulations. For the circumbinary and flyby examples, we find that the APR calculations are accurate to $< 0.1\%$. For the planet-disc interaction example we measure both the mass and the radial location of the planet and find that it is accurate to at least $9\%$ (but is generally more accurate than this).
    \item \emph{Fast:} Every APR simulation showed a speed up, offering between $1.07-6.62\times$ than when compared to a simulation with the same resolution but globally. The speed up is application-dependent, with the flyby example being the most rapid and the circumbinary disc the least amenable to speedup in this manner.
    \item \emph{Uses less storage:} Because our APR simulations use fewer particles in total, they require between $15 - 27\%$ of the storage of an equivalent globally resolved calculation.
\end{enumerate}

Our example applications suggested optimal sizes of the refinement region as a guide for future calculations. We found accuracy of the implementation depends sensitively on whether or not key features were uniformly resolved. We also demonstrated that the location of the refinement region can be dynamic and note that derefinement is possible. Finally we showed that APR can increase the resolution of simulations at low cost; for the flyby example we achieved a local resolution of $32$ million particles with totals of either $3.0$ or $5.7$ million particles.

We showed examples limited to hydrodynamical simulations, but \textsc{Phantom} also includes dust \citep{Laibe:2012vu,Laibe:2014ms}, magnetohydrodynamics \citep{Tricco:2012fk}, self-gravity, general relativity \citep{Liptai:2019vg} and can be coupled with \textsc{MCFOST} \citep{Pinte:2006nw,Pinte:2009ye,Nealon:2020bq}. For brevity we leave dedicated testing of our APR implementation with these features to future works.

\begin{acknowledgement}
The authors thank James Wurster for significant contributions in the early stages of this work and the referee for helpful suggestions. RN acknowledges Matthew Bate, Richard Booth, Terry Tricco and Alessia Franchini for discussions and Sahl Rowther for aesthetics assistance.
\end{acknowledgement}

\paragraph{Funding Statement}
R.N. acknowledges support from UKRI/EPSRC through a Stephen Hawking Fellowship (EP/T017287/1). This work was performed using Avon, the HPC clusters at the University of Warwick. DP is grateful for Australian Research Council funding via DP220103767 and DP240103290. We thank the Monash-Warwick alliance for sponsorship of the {\sc phantom} users workshop where this work commenced and Nicol\'{a}s Cuello for his hospitality.

\paragraph{Data Availability Statement}
The data to reproduce figures in this study are openly available on Zenodo at http://doi.org/10.5281/zenodo.11581005. All data from the simulations will be made available upon request to the corresponding author. The software used to create and visualise the simulations are publicly available:\\
\newline
\begin{tabular}{ll}
\textsc{Phantom}: & \url{https://github.com/danieljprice/phantom} \\
                  & \citep{Phantom} \\
\textsc{Sarracen}: & \url{https://github.com/ttricco/sarracen} \\
                   & \citep{Sarracen}
\end{tabular}

\printendnotes

\printbibliography

\appendix
\section{Simple tests}
\label{section:boxtests}
To demonstrate some of the key choices of our implementation we use the straightforward example of a sound wave in a periodic box simulated with \textsc{Phantom}. The density is set by
\begin{align}
    \rho(x) = \rho_0 + A\sin\left(\frac{2\pi x}{L}\right),
\end{align}
where $A = 0.02$ is the amplitude of the perturbation and $\rho_0 = 1$. The thermal energy is perturbed in the same manner with the same amplitude $A$ but scaled by $P_0/\rho_0$, the unperturbed pressure and density. The fluid has an unperturbed sound speed of $c_{{\rm s},0}=1.0$ and we adopt an adiabatic equation of state. Unless otherwise stated the velocity is set to $v_x=0.001$, $v_y=v_z=0$. The particles are initially set on a close-packed lattice with the above density distribution. We add noise to the lattice by randomly adjusting the $x$ and $y$ positions of the particles by $0.0001 \times$ the length of the box. We use 64 particles across the box of width $dx = 1.0$ and $dy=1.0$, corresponding to a total of $N=56832$ particles initially.

The refinement region is a circle centred at $x=0.5$, $y=0.5$ with a radius of $r=0.10$. In contrast to our example applications, these simulations only make use of one refinement level ($\ell=1$) which corresponds to a doubling of the particle number inside this region. The first split occurs at $t=0.5$ to allow the particles to settle a bit first and the simulation is then performed until $t=2.0$, corresponding to five wave-crossings in total. After the split the simulation has around $N=63500$ particles.

Our results in this section are all shown with the cross section of the density, $\rho(x)$, across the box divided by the initial average density, $\overline{\rho_{\rm IC}}$, to avoid small variations in the background density caused by kernel bias. For clarity we also restrict the visualisation to only include particles between $0.4 < y < 0.6$, essentially making it a cross section centred on $y=0.5$. On each plot we also display the average density as a function of position across the box for both the APR simulation and a reference simulation without any refinement. This allows us to easily compare how much noise is introduced as a result of the splitting/merging and how accurate these simulations are compared to the simulations that do not include APR. Movies of some of the simulations in this section are available \href{https://www.youtube.com/playlist?list=PLVCh7-wtOgbJHwf-1J9HX4i9E4pFU5aUF}{online}.

\subsection{Density discontinuity - the `blip'}
A common feature in all of our simulations is a density discontinuity at the refinement boundary (i.e. `the blip' between the different resolution regions). It appears as a slight increase in the density just outside the refinement region and a slight decrease immediately inside. This discontinuity does not introduce spurious waves, has a constant amplitude and does not appear to affect the density profile inside the refinement region. Its apparent prominence is because of our use of a comparatively small amplitude sound wave as our test case - we chose it specifically so that any spurious features would be easily identifiable. This feature maintains a constant amplitude (here, of $<5\%$ of the original density) and in our examples in Section~\ref{section:examples} is barely visible in renderings (e.g. Figure~\ref{fig:planets_rendered}), similar to the density profiles seen in examples with GIZMO \citep{Angles-Alcazar:2021jj,Franchini:2022nh,Duffel:2024ng}. Importantly, the blip does not seem to affect the propagation of the wave itself.

This density blip is a manifestation of a known issue in APR, where the splitting and merging procedure introduces small errors into the density distribution. \citet{FeldmanBonet:2007bh} formally established that an error in the density distribution is introduced each time a parent particle is replaced by children particles when the children all have the same mass. When a split occurs and the children particles are introduced, they should be placed symmetrically around the parent to ensure conservation of angular momentum \citep[e.g.][]{Lopez:2013nj}. However, the introduction of these new particles also introduces this error, the magnitude of which is a function of the distance the children are placed at $r_{\rm sep}$. In turn, this means the error can be minimised by adjusting the locations of the children.

\citet{Lopez:2013nj} extended this idea to consider the error introduced in the derivative of the density, suggesting that this was the more relevant error to minimise for subsequent calculations. Further works looked at how best to minimise this error that is introduced every time particles are split or merged with various $r_{\rm sep}$ \citep[e.g.][]{Lopez:2013nj}, particle shuffling/relaxing \citep[e.g.][]{Lind:2012bj,Diehl:2015vu,Sun:2017jo}, the use of ghost particles and blending techniques \citep[e.g.][]{Barcarolo:2014vu,Chiron:2018hq,Gao:2022wd} to mitigate this at the boundaries of refinement regions. 
\citet{Barcarolo:2014vu} eliminated this density blip by employing a blending zone between each refinement level, where both children and parents co-exist but their contribution to the density summation is graduated by their distance across the blending zone. This approach was also adopted by \citet{Gao:2022wd} but only as new particles moved into the higher refinement level, allowing them to regularise first. While promising, blending is difficult to implement in compressible flows due to mass conservation in the blending zone.

The alternative is particle relaxation (shuffling): to minimise the error introduced when particles were split, \citet{Diehl:2015vu} used a WVT shuffling method to rearrange the particles to a lower error state compared to the original parent distribution. \citet{Yang:2019vg} employed multiple, stepped refinement regions to control the ratio of the smoothing lengths across the boundaries and mitigate instabilities \citep[similar to ][]{Borve:2001oe}. In our implementation we employ relaxation, but find that it is only important when a large number of particles are split or merged at once.

\begin{figure}[t]
    \centering
    \includegraphics[width=\textwidth]{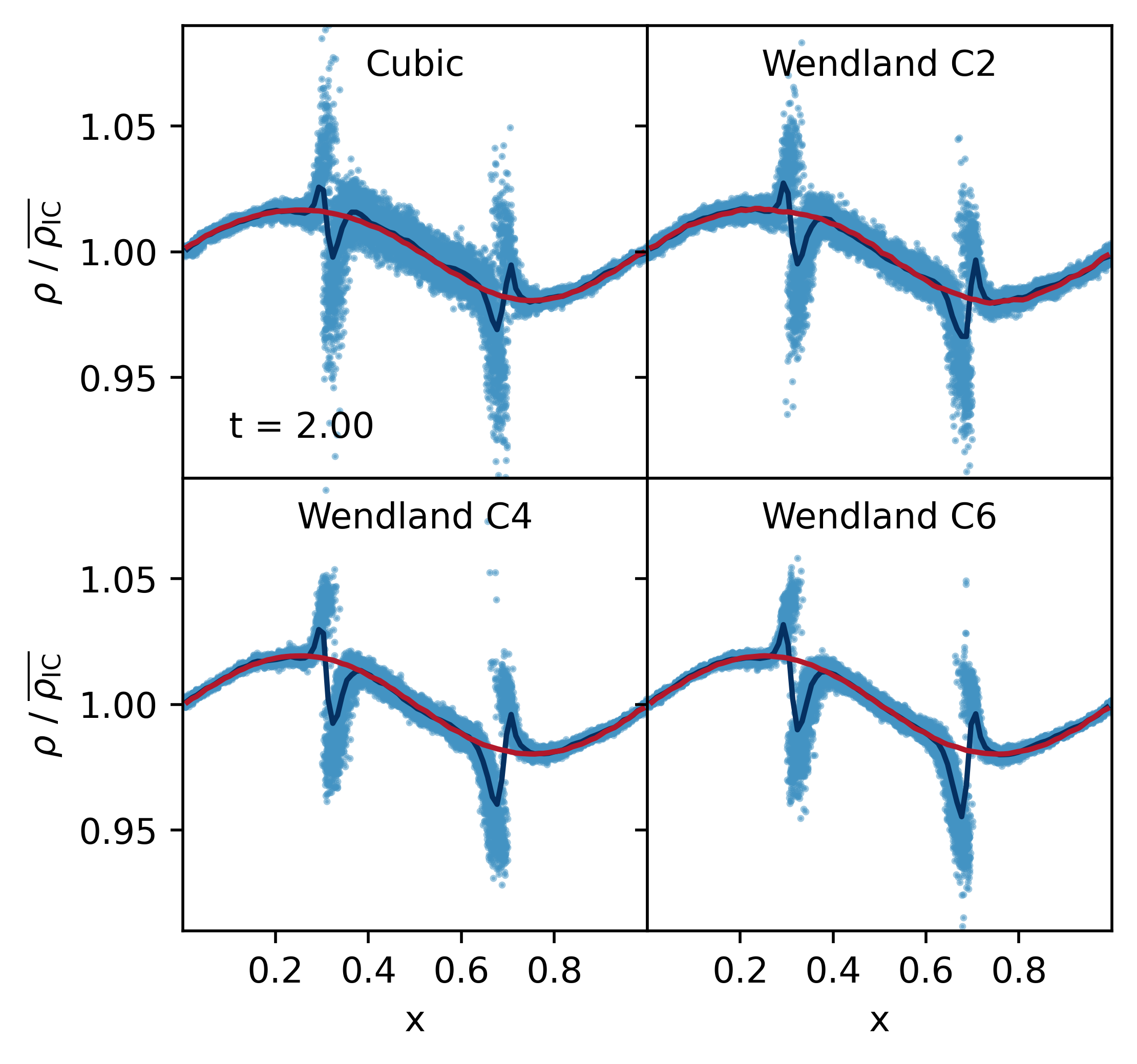}
    \caption{Wave in a box test showing the effect of different kernels with APR. The blue points show the particles within the $y=0.5$ cross-section across the $x$ dimension of the box at $t=2.0$. The dark blue line shows their average density, while the red line shows the average density of the same simulation without APR. The density discontinuity at the refinement boundary maintains a constant amplitude across all of the simulations.}
    \label{fig:box_kernels}
\end{figure}

\begin{figure*}[t]
    \centering
    \includegraphics[width=0.8\textwidth]{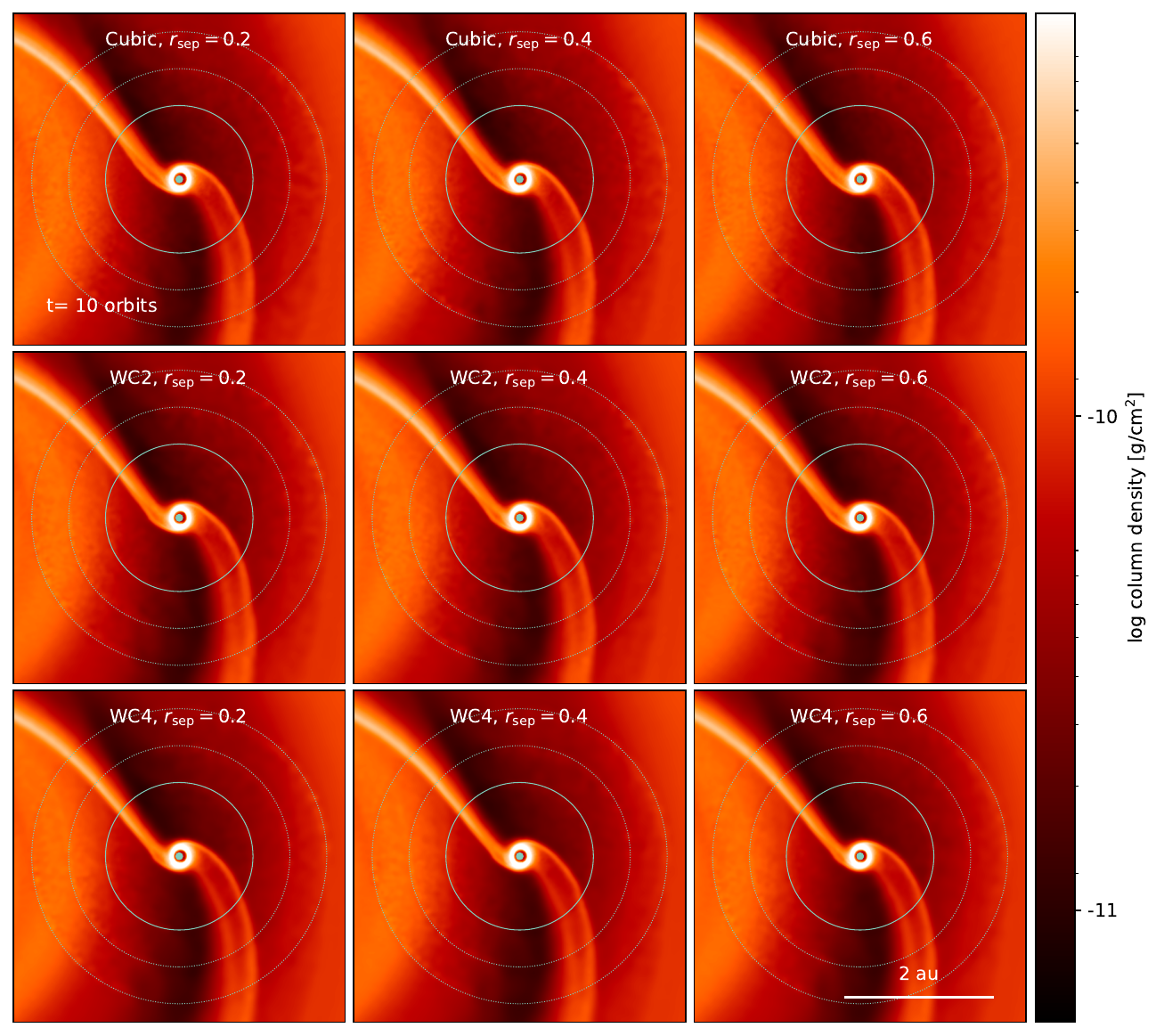}
    \caption{Planet disc interaction test showing the effect of different kernel choices and $r_{\rm sep}$ on the noise introduced at the refinement boundary in a full simulation. Refinement boundaries are indicated in the same way as Figure~\ref{fig:HD142527_rendered}. The boundary is smoothest for the Wendland C4 kernel when $r_{\rm sep} = 0.2$ but we note the difference is marginal.}
    \label{fig:planet_disc_separation}
\end{figure*}

\begin{figure}[h!]
    \centering
    \includegraphics[width=\textwidth]{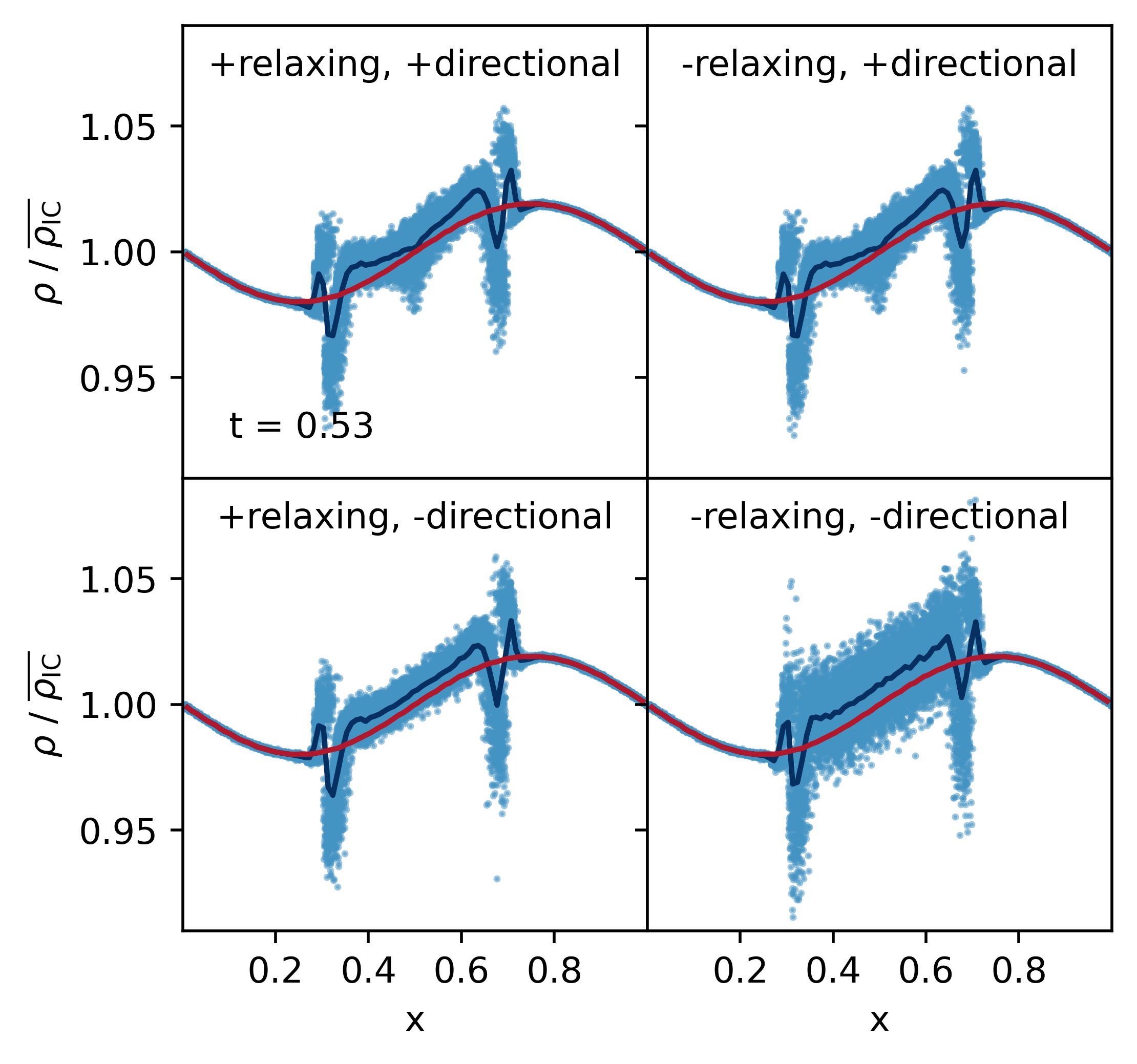}
    \includegraphics[width=\textwidth]{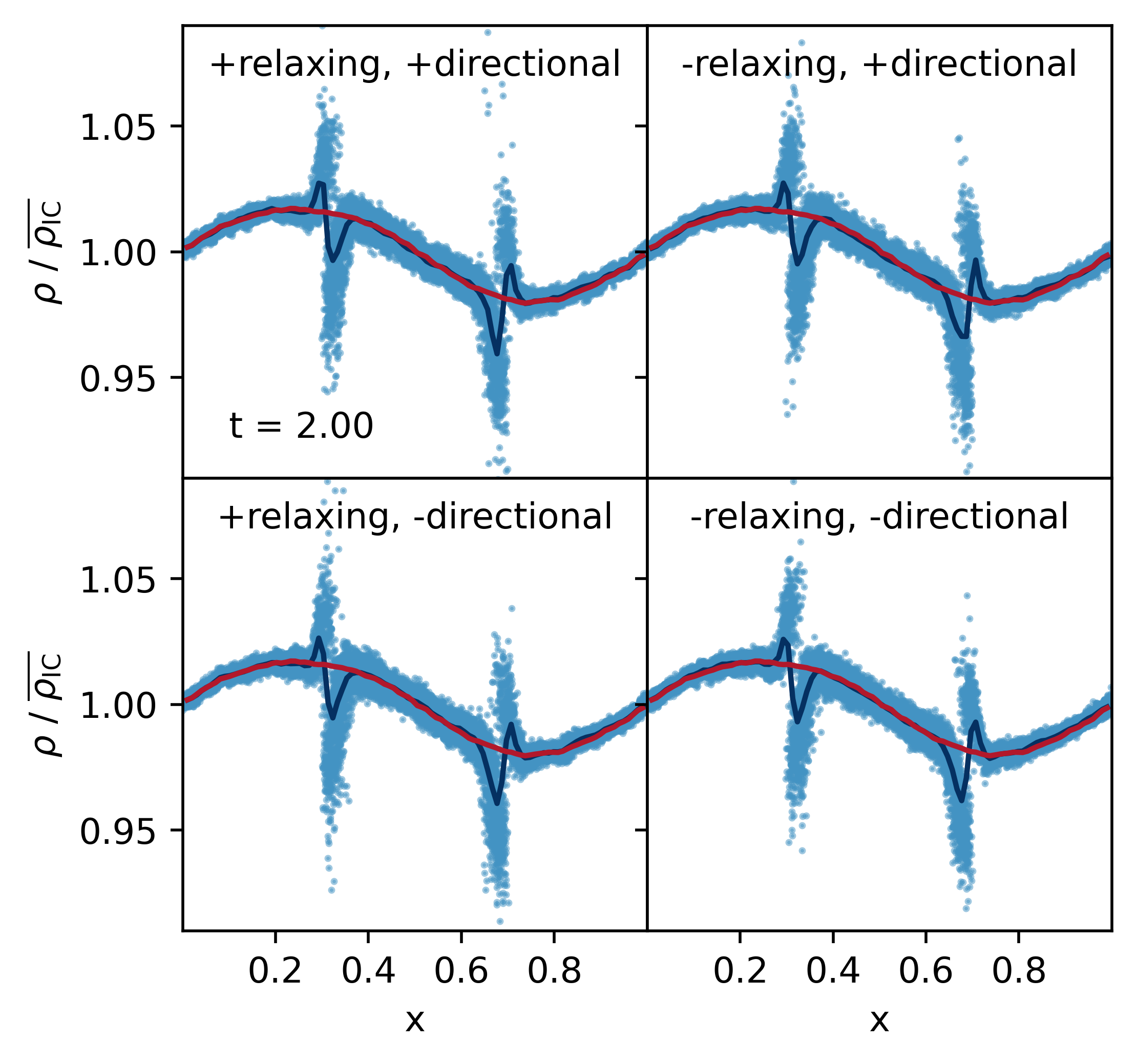}
    \caption{Wave in a box test showing the effect of different particle placement options when a split or merge occurs shown at the initial split (upper, $t=0.5$) and at the end of the simulation (lower, $t=2.0$). The colours are the same as in Figure~\ref{fig:box_kernels}. Relaxing is most successful when the split first occurs but makes negligible difference during the course of the simulation. Directional splitting also makes little difference in the long term but does prevent particles from splitting across the boundary.}
    \label{fig:box_relaxing}
\end{figure}

\subsection{Kernel choice}
Figure~\ref{fig:box_kernels} shows the final snapshot of our test simulation comparing the use of the Cubic spline \citep{Monaghan:1985jl}, Wendland C2, Wendland C4 and Wendland C6 kernels \citep{Wendland:1995hd}, summarised in Table~\ref{tab:kernels}. The density profile in the refinement region (blue line) is in excellent agreement with the expected profile (red line) for all kernel choices. As might be expected, the kernels with more neighbours have less noise inside the refinement region and at the boundary. In particular, the cubic has the largest noise at the refinement region boundary, the largest spread in density inside the refinement region and has the least accurate density profile (although the difference here is marginal).

Figure~\ref{fig:planet_disc_separation} shows our planet disc interaction test at $t=10$ orbits but with different combinations of $r_{\rm sep}$ and either a Cubic, Wendland C2 or Wendland C4 kernel. Here we have zoomed in to see the boundaries of the refinement zone to examine the effect of these choices. Moving between kernels, the WC4 in the lowest row is smoothest while the cubic has the largest noise at the boundaries. For any of the kernels, increasing $r_{\rm sep}$ beyond 0.2 also corresponds to an increase in the noise at the boundary (seen most clearly in the WC4 case). Both of these tests confirm that a higher order kernel is more effective at mitigating the density `blip' but that the difference is marginal. For our example applications we adopt the Wendland C2 kernel as a balance between accuracy and computational expense.

\begin{table}
    \centering
    \begin{tabular}{ccc}
        Kernel & $N_{\rm neigh}$ & $h_{\rm fact}$\\
        M4 Cubic & 57.9 & 1.2\\
        Wendland C2 & 92 & 1.3\\
        Wendland C4 & 137 & 1.5\\
        Wendland C6 & 356 & 1.6\\
    \end{tabular}
    \caption{Summary of the $h_{\rm fact}$ employed and the average number of neighbours $N_{\rm neigh}$ for the kernels tested in Figure~\ref{fig:box_kernels} \citep[e.g.][]{Phantom}.}
    \label{tab:kernels}
\end{table}

\subsection{Adjusting the split}
\label{section:adjusting_split}
Figure~\ref{fig:box_relaxing} demonstrates the effectiveness of both particle relaxing (as per Section~\ref{section:relax}) and directional splitting, where particles are split tangentially to the refinement boundary.
Importantly, once the particles have settled after their initial split (Figure~\ref{fig:box_relaxing}, top panel) there is no significant difference between the methods in the density profiles. However, their time to compute is quite different with the simulations that include relaxing taking much longer. In practice, we thus adopt directional splitting as it is fast, gives an improved initial density estimate and prevents particles from being immediately placed across the boundary into a low resolution region when they are split. To improve computational time by default we only employ the relaxing routine when refinement regions are activated and there is a large number of particles that are split or merged at once.

\begin{figure}
    \centering
    \includegraphics[width=\textwidth]{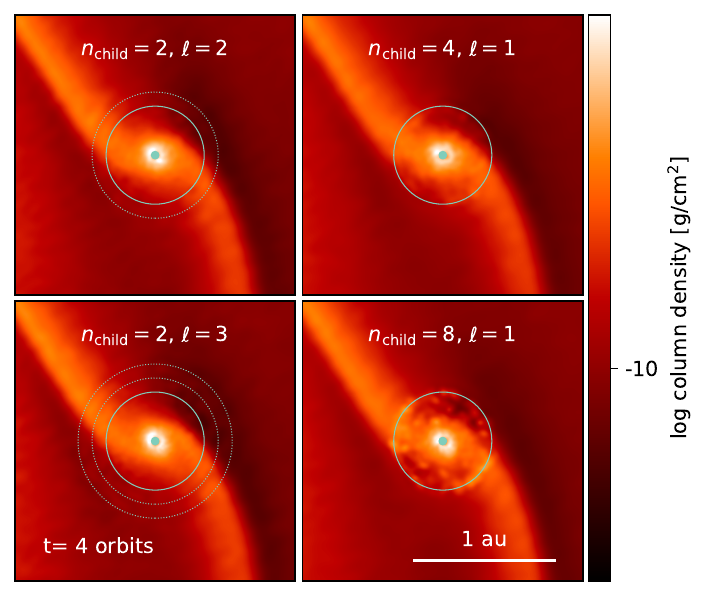}
    \caption{
    The planet disc interaction test, examining the effect of having different $n_{\rm child}$. The left column has $n_{\rm child}=2$ with nested refinement levels, the upper right has $n_{\rm child}=4$ and one level of refinement, the lower right $n_{\rm child}=8$ and one level. The refinement zones are shown as in Figure~\ref{fig:HD142527_rendered}. Rows have the same local resolution around the planet. While the nested refinement zones do add noise at the boundaries, this is demonstrably less than bigger numbers of children.}
    \label{fig:nchild_planet_test}
\end{figure}

\subsection{Number of children}
\label{section:four_is_too_many}
In our tests so far we have used $n_{\rm child} =2$, splitting parents into two children and merging two children to become a parent. In Figure~\ref{fig:nchild_planet_test} we show the impact of this choice by considering $n_{\rm child} = 4$ and $n_{\rm child} = 8$. We compare this to simulations with $n_{\rm child} = 2$ and $\ell = 2,3$ respectively which have equivalent local resolutions.

To split one parent into four children we follow the method outlined in Section~\ref{section:split} and treat those two children as the diagonal corners of a square. We then add an additional two particles on the opposing diagonal which ensures that the face of the square of four children particles is parallel to the refinement boundary. For eight children we create two squares that are offset by $0.1 \times$ the smoothing length of the original parent, with the centre of mass of the resultant cube centred on the original parent. For our merging routines we simply edit our modified $k$-d tree to return cells with $n_{\rm child} = 4$ and $n_{\rm child} = 8$.

Figure~\ref{fig:nchild_planet_test} shows the planet disc interaction test at $t=4$ orbits for these choices, where the centrally refined zone has the same width ($r_{\ell}=0.35$au and the steps into the refinement zone at increments of $0.10$au). The noise introduced by the nested zones is visibly lower in both cases; in the $n_{\rm child}=8$ case the noise is the largest and is also prominent across the whole refined region, even after it has had the opportunity to settle after the first split. We note that while this test may be improved by relaxing at each step, in practice this becomes computationally extremely expensive. This implies that using a smaller number of children --- even if it means having nested refinement regions --- is the preferred approach. In other words, even four children is too many.

\subsection{Size of the APR zone}
\label{section:aprzone_size}

We test the effect of the size of the refinement region using the jet test presented in \citet{Price:2024mj}, Appendix E, Figure 14. In this test a jet of gas shoots in one direction from a fixed location in an empty domain, the gas flares out as it expands slightly. Based on the results of \citet{Price:2024mj} we set the Mach number of the gas to 10 to ensure a fairly narrow jet. The particles are injected as a cylinder with radius $R=1$ 16 particles in each layer. The rest of the parameters in this simulation are scale free.

We apply four different APR zones to this jet test with $\ell=1$ for all but $r_{\ell} = 3, 5, 10$ and $15$. The refinement boundary of each different zone has the same position of $y=5$ so that particles are always split at the same location but they are merged at different distances from the launching point of the jet. Figure~\ref{fig:firehose} shows this with the refinement zone indicated with a green dashed circle. We chose an injection velocity of $v_y=37.5$ and calculate how many sound crossing times the particles will experience between refinement and derefinement to be $<1$, $1.3$, $2.7$ and $4.0$.

Figure~\ref{fig:firehose} shows the column density of the jets at $t=50$ with the refinement zones superimposed. For the tests where particles have fewer sound crossing times we find that there is a slight narrowing of the jet within the refinement zone but for the final test where particles have several sound crossing times this narrowing is not apparent. This test demonstrates that particles require several sound crossing times between the refinement boundaries to allow them to relax before they are derefined.

\begin{figure}[h!]
    \centering
    \includegraphics[width=\textwidth]{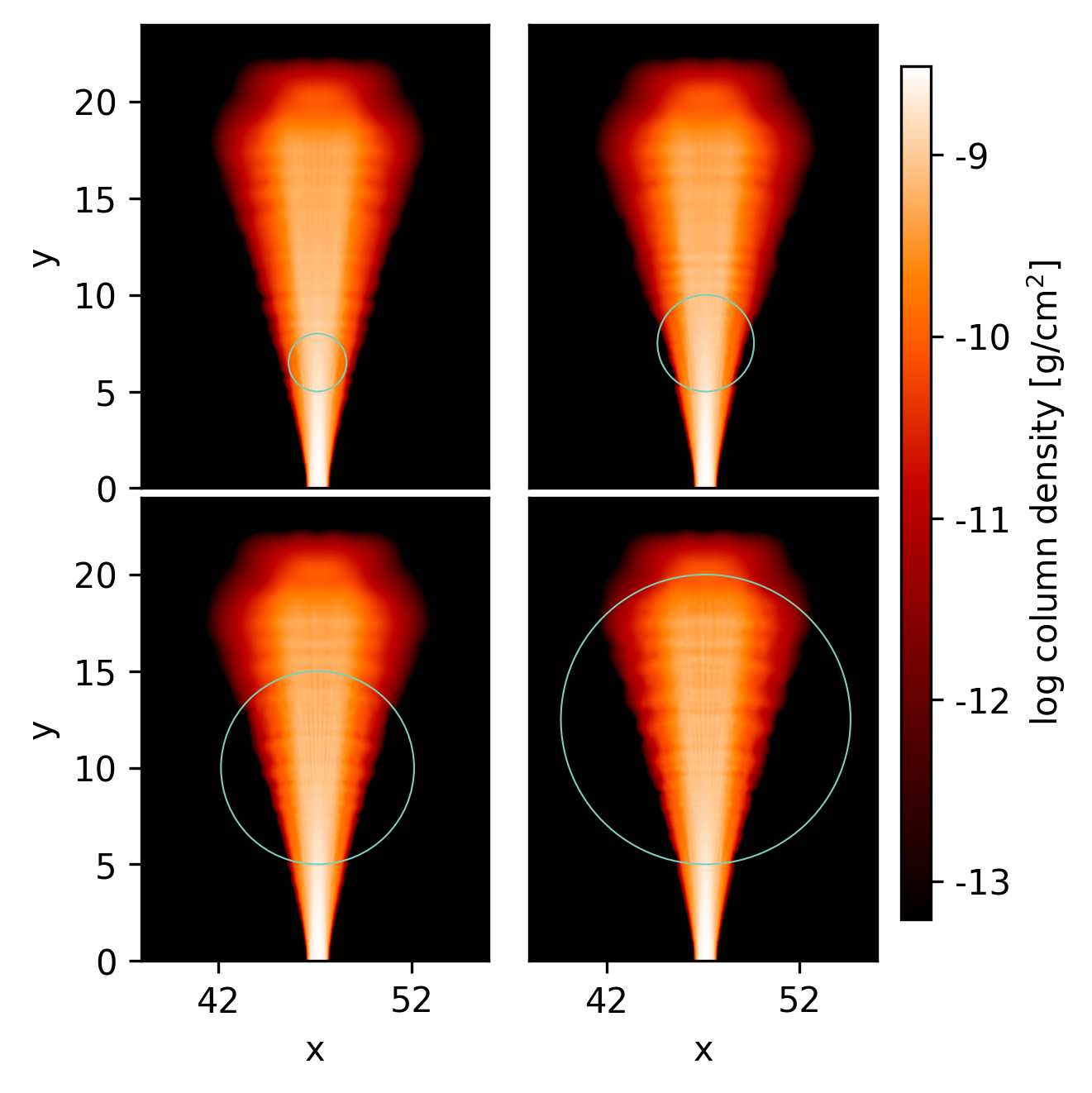}
    \caption{Testing the size of the refinement zone, where the four simulations are characterised by the number of sound crossings that can occur between the refinement and derefinement boundaries. The refinement region is indicated with a green circle in the figure.}
    \label{fig:firehose}
\end{figure}

\end{document}